\documentclass[a4paper,12pt]{article}

\usepackage{hyperref}
\hypersetup{
    colorlinks=true,
    linkcolor=blue,
    filecolor=magenta,      
    urlcolor=cyan,
    citecolor=green
    }

\usepackage{latexsym}
\usepackage{a4wide}
\usepackage{graphicx, subfigure} 
\usepackage{cite} 
\usepackage{tabularx}
\usepackage{array}
\usepackage{graphics}     
\usepackage{amsmath}
\usepackage{amssymb}
\usepackage{longtable} 
\usepackage{verbatim} 
\usepackage[table]{xcolor}
\usepackage{booktabs}
\usepackage{hyperref} 
\usepackage[utf8]{inputenc}
\usepackage{multirow}
\usepackage{soul}

\newcolumntype{C}[1]{>{\centering\arraybackslash}p{#1}}
\newcolumntype{L}[1]{>{\raggedright\arraybackslash}p{#1}}
\newcolumntype{R}[1]{>{\raggedleft\arraybackslash}p{#1}}

\definecolor{dgreen}{rgb}{0.0, 0.5, 0.0}

\begin{document}

\hfill {\tt CERN-TH-2023-182, MITP-23-54}  

\def\thefootnote{\fnsymbol{footnote}}
 
\begin{center}

\vspace{2.cm}

{\Large\bf {\emph{B} anomalies in the post $\boldsymbol{R_{K^{(*)}}}$ era}}

\setlength{\textwidth}{11cm}
                    
\vspace{2.cm}

{\large\bf  
T.~Hurth$^{a}$,\,
F.~Mahmoudi$^{b,c,d}$,\,
S.~Neshatpour$^{b,e}$
}
 
\vspace{0.5cm}
{\em $^a$PRISMA+ Cluster of Excellence and  Institute for Physics (THEP),\\
Johannes Gutenberg University, D-55099 Mainz, Germany}\\[0.2cm]
{\em $^b$Universit\'e de Lyon, Universit\'e Claude Bernard Lyon 1, CNRS/IN2P3, \\
  Institut de Physique des 2 Infinis de Lyon, UMR 5822, F-69622, Villeurbanne, France}\\[0.2cm]
{\em $^c$CERN, Theoretical Physics Department, CH-1211 Geneva 23, Switzerland,}\\[0.2cm]
{\em $^d$Institut Universitaire de France (IUF),}\\[0.2cm]
{\em $^e$INFN-Sezione di Napoli, Complesso Universitario di Monte S. Angelo,\\ Via Cintia Edificio 6, 80126 Napoli, Italy}

\end{center}

\renewcommand{\thefootnote}{\arabic{footnote}}
\setcounter{footnote}{0}

\vspace{1.cm}
\thispagestyle{empty}
\centerline{\bf ABSTRACT}
\vspace{0.5cm}
We discuss the status of $b \to s \ell^+\ell^-$ decays in the post-$R_{K^{(*)}}$ era. 
The recent LHCb update of $R_K$ and $R_{K^*}$ measurements
which are now compatible with the Standard Model, constrain new physics contributions to be lepton flavor universal, allowing only small deviations from this limit.
Besides the latest LHCb measurements of $R_K$ and $R_{K^{*}}$, we also include the recent CMS measurements of $R_K$ and of the branching ratio of $B^+ \to K^+ \mu^+\mu^-$.
We present a model-independent analysis of the $b\to s \ell^+ \ell^-$ data and investigate the implications of the different sets of observables. In addition, we consider multi-dimensional fits and discuss the significance of more complex new physics scenarios compared to one- and two-dimensional scenarios.

\newpage

\section{Introduction}
Over the last nine years the LHCb collaboration reported hints for lepton non-universality at the  3$\sigma$ level via measurements of the ratios $R_K$ and $R_{K^{*}}$  (see~\cite{LHCb:2017avl,LHCb:2021trn}
and references therein). 
The $R_K$ and $R_{K^*}$ which are defined as the ratios of the branching fractions of $B\to K^{(*)}\ell^+\ell^-$ for muons vs. electrons are theoretically very clean, with uncertainties of less than one percent and central values close to unity in the Standard Model (SM) due to the universality of the lepton flavours~\cite{Hiller:2003js,Bordone:2016gaq}.
In addition, there are long-standing tensions in the angular observables and branching ratios of exclusive $b \to s$ observables~\cite{LHCb:2013ghj,LHCb:2014cxe,LHCb:2015tgy,LHCb:2015wdu,LHCb:2015svh,LHCb:2020lmf,LHCb:2020gog,LHCb:2021xxq,LHCb:2021zwz}.
{In general the observables of the exclusive decays are dependent on local matrix elements (form factors), as well as non-local ones which often make it difficult to distinguish between possible new physics effects and hadronic effects. Although some of the angular observables are less sensitive to the form factors, they do depend on non-local hadronic contributions, which are not well known. The significance of the anomalies in exclusive decays is therefore dependent on the estimated size of the non-local effects. 
Recent theoretical progress in  the evaluation of the non-local contributions~\cite{Bobeth:2017vxj,Gubernari:2020eft,Gubernari:2022hxn} indicate that the non-factorisable power corrections are small.}
The crucial point of the previous situation was that the deviations in the theoretically clean ratios on one side and in the angular observables and branching ratios on the other side could be consistently described with the same new physics {scenarios}. This consistency was again increased
with the updated measurement of BR($B_s\to \mu^+\mu^-$) from last year~\cite{CMS:2022mgd}. The combination of this result with the ATLAS and LHCb measurements~\cite{ATLAS:2018cur,LHCb:2021awg,LHCb:2021vsc}, BR$(B_s \to \mu^+ \mu^-)_{\rm exp}^{\rm comb.} = \left(3.52^{+0.32}_{-0.30} \right) \times 10^{-9}$ as given in~\cite{Neshatpour:2022pvg} is in agreement with the SM within $1\sigma$, suppressing large new physics contributions in the Wilson coefficient~$C_{10}$.%
However, the LHCb collaboration recently presented new measurements of the ratios which turn out to be compatible with the Standard Model~\cite{LHCb:2022vje} 
{
\begin{align*}\label{eq:LHCb_RK_updated}
\footnotesize
    \begin{cases}
        R_{K}([0.1-1.1])   = 0.994~^{+0.090}_{-0.082}~^{+0.029}_{-0.027}\,, \\[6pt]
        R_{K}([1.1-6.0])  = 0.949~^{+0.042}_{-0.041}~^{+0.022}_{-0.022}\,, \\
   \end{cases} 
   \qquad \footnotesize
   \begin{cases}
        R_{K^{*}}([0.1-1.1]) = 0.927~^{+0.093}_{-0.087}~^{+0.036}_{-0.035}\,, \\[6pt]
        R_{K^{*}}([1.1-6.0]) =1.027~^{+0.072}_{-0.068}~^{+0.027}_{-0.026}\,.
   \end{cases} 
\end{align*}
}

In this paper, we analyse the current situation in a model-independent way. Clearly, the tensions in the angular observables and branching ratios are untouched by the new LHCb measurements.  We analyse the two sets of $b \to s$ data 
separately, namely the theoretically clean ratios together with  BR($B_{s,d}\to \ell^+\ell^-$) on one side and the angular observables and branching ratios on the other side.

We also include the very recent measurements of $R_K$ and the branching ratio of $B^+ \to K^+ \mu^+\mu^-$ by the CMS collaboration~\cite{CMS:2023klk}, which both turn out to be compatible with the SM predictions. In addition, we update the CKM parameters where we  have updated the PDG~2020~\cite{ParticleDataGroup:2020ssz} values to PDG~2022~\cite{ParticleDataGroup:2022pth}, with the old and new inputs given below
\begin{table}[h!]
\centering
\scalebox{0.85}{
{
\begin{tabular}{c|c|c|c|c}
& $\lambda$ & $A$ & $\bar{\rho} $ & $\bar{\eta}$ \\\hline
PDG (2020) & $0.22650 \pm 0.00048$ & $0.790^{+0.017}_{-0.012}$ & $0.141^{+0.016}_{-0.017}$ & $0.357\pm 0.011$ \\[4pt]
PDG (2022) & $0.22500 \pm 0.00067$ & $0.826^{+0.018}_{-0.015}$ & $0.159\pm0.010$ & $0.348\pm0.010$ 
\end{tabular}}
}
\end{table}

The complete list of the observables used in the present fits can be read off the corresponding list in our previous analysis in Refs.~\cite{Hurth:2021nsi,Neshatpour:2022pvg}. {For our analysis we have used the \texttt{SuperIso} public program~\cite{Mahmoudi:2007vz,Mahmoudi:2008tp,Mahmoudi:2009zz,Neshatpour:2021nbn,Neshatpour:2022fak} assuming 10\% uncertainty for the unknown non-factorisable power corrections (see Ref.~\cite{Hurth:2016fbr} for more details).} 
{For other global analyses with the  updated LHCb measurement of $R_{K^{(*)}}$ (not including the recent CMS measurement) see for example~\cite{Alguero:2023jeh,Ciuchini:2022wbq,Greljo:2022jac}.}

This paper is organised as follow{, in the next section we show the one- and two-dimensional fits for different sets of observables. In section~\ref{sec:clean} we consider clean observables and discuss the impact of new LHCb measurement for the ratios, and in section~\ref{sec:all_except_clean} the fit to the rest of the observables are given where the impact from the CMS measurement on BR($B^+ \to K^+ \mu^+ \mu^-$) as well as the updated CKM values are visible. In section~\ref{sec:all_obs} the fit to all $b\to s$ data are given and the impact of various sets of observables are discussed. Section~\ref{sec:global} includes a 12-dimensional fit and shows via the Wilks' test that beyond $C_9$ adding further degrees of freedom only improves the fit marginally. Finally, we summarise in  section~\ref{sec:summary}.}

\section{One- and two-dimensional fits}
\subsection{Fits to clean \texorpdfstring{$b \to s\ell\ell$}{b->sll} observables}\label{sec:clean}
\begin{table}[th!]
\begin{center}
\setlength\extrarowheight{0pt}
\scalebox{0.99}{
\begin{tabular}{|l|r|r|c|}
\hline 
\multicolumn{4}{|c|}{\footnotesize { Only LFUV ratios and $B_{s,d}\to \ell^+ \ell^-$} \vspace{-0.1cm}}\\	 									
 \multicolumn{4}{|c|}{\small {{\bf pre-$\boldsymbol{R_{K^{(*)}}}$ update}} \;($\chi^2_{\rm SM}=	30.63	$)}\\ \hline								
& b.f. value & $\chi^2_{\rm min}$ & ${\rm Pull}_{\rm SM}$  \\										
\hline \hline										
$\delta C_{9}^{e} $    	& $ 	0.83	\pm	0.21	 $ & $ 	10.8	 $ & $	4.4	\sigma	 $  \\
$\delta C_{9}^{\mu} $    	& $ 	-0.80	\pm	0.21	 $ & $ 	11.8	 $ & $	4.3	\sigma	 $  \\
\hline										
$\delta C_{10}^{e} $    	& $ 	-0.81	\pm	0.19	 $ & $ 	8.7	 $ & $	4.7	\sigma	 $  \\
$\delta C_{10}^{\mu} $    	& $ 	0.50	\pm	0.14	 $ & $ 	16.2	 $ & $	3.8	\sigma	 $  \\
\hline							          			
\hline										
$\delta C_{\rm LL}^e$	& $ 	0.43	\pm	0.11	 $ & $ 	9.7	 $ & $	4.6	\sigma	 $  \\
$\delta C_{\rm LL}^\mu$    	& $ 	-0.33	\pm	0.08	 $ & $ 	12.4	 $ & $	4.3	\sigma	 $  \\
\hline										
\end{tabular}
}
\qquad
{
\begin{tabular}{|l|r|r|c|}
\hline 
\multicolumn{4}{|c|}{\footnotesize {Only LFUV ratios and $B_{s,d}\to \ell^+ \ell^-$} \vspace{-0.1cm}}\\	 									
 \multicolumn{4}{|c|}{\small {{\bf post-$\boldsymbol{R_{K^{(*)}}}$ update}} \;($\chi^2_{\rm SM}=	9.37	$)}\\ \hline								
& b.f. value & $\chi^2_{\rm min}$ & ${\rm Pull}_{\rm SM}$  \\										
\hline \hline										
$\delta C_{9}^{e} $    	& $ 	0.17	\pm	0.16	 $ & $ 	8.2	 $ & $	1.1	\sigma	 $  \\
$\delta C_{9}^{\mu} $    	& $ 	-0.18	\pm	0.16	 $ & $ 	8.1	 $ & $	1.1	\sigma	 $  \\
\hline										
$\delta C_{10}^{e} $    	& $ 	-0.15	\pm	0.14	 $ & $ 	8.3	 $ & $	1.1	\sigma	 $  \\
$\delta C_{10}^{\mu} $    	& $ 	0.15	\pm	0.12	 $ & $ 	7.7	 $ & $	1.3	\sigma	 $  \\
\hline							          			
\hline										
$\delta C_{\rm LL}^e$	& $ 	0.08	\pm	0.08	 $ & $ 	8.2	 $ & $	1.1	\sigma	 $  \\
$\delta C_{\rm LL}^\mu$    	& $ 	-0.09	\pm	0.07	 $ & $ 	7.7	 $ & $	1.3	\sigma	 $  \\
\hline										
\end{tabular}
}
\caption{{One operator NP fit to clean observables  before and after update of $R_{K^{(*)}}$ by the LHCb  collaboration.}
\label{tab:2022_2023_1D_clean}} 
\end{center} 
\end{table}
First, we analyse the significance of new physics (NP) within the clean observables, $R_{K^{(*)}}$ and BR($B_{s,d} \to \mu^+ \mu^-$). 
In Table~\ref{tab:2022_2023_1D_clean} we show   the  one-operator fits to these clean observables, both before\footnote{{In this paper, pre-$R_{K^{(*)}}$ indicates the fit to the data before the LHCb update on $R_{K^{(*)}}$ as given in~\cite{Neshatpour:2022pvg}}.} and after the latest $R_{K^{(*)}}$ measurements. The change is a drastic one, as only small deviations from lepton-universality are now allowed. There are still lepton flavour universality violating (LFUV) ratios, namely  $R_{K^0_S}^{\rm LHCb}([1.1-6.0])$, $R_{K^{*+}}^{\rm LHCb}([0.045-6.0])$~\cite{LHCb:2021lvy} and $R_K^{\rm LHCb}([1.1-6.0])$~\cite{LHCb:2022vje}
with 1.7, 1.4 and $1.1\sigma$ NP significance, respectively.\footnote{{A re-analysis of $R_{K^0_S}^{\rm LHCb}([1.1-6.0])$ and $R_{K^{*+}}^{\rm LHCb}([0.045-6.0])$} regrading possible misidentifications  would not change the NP significances  much  given the large experimental uncertainties~\cite{PO:pricom}.}
\begin{figure}[t!]
\centering
\includegraphics[width=0.45\textwidth]{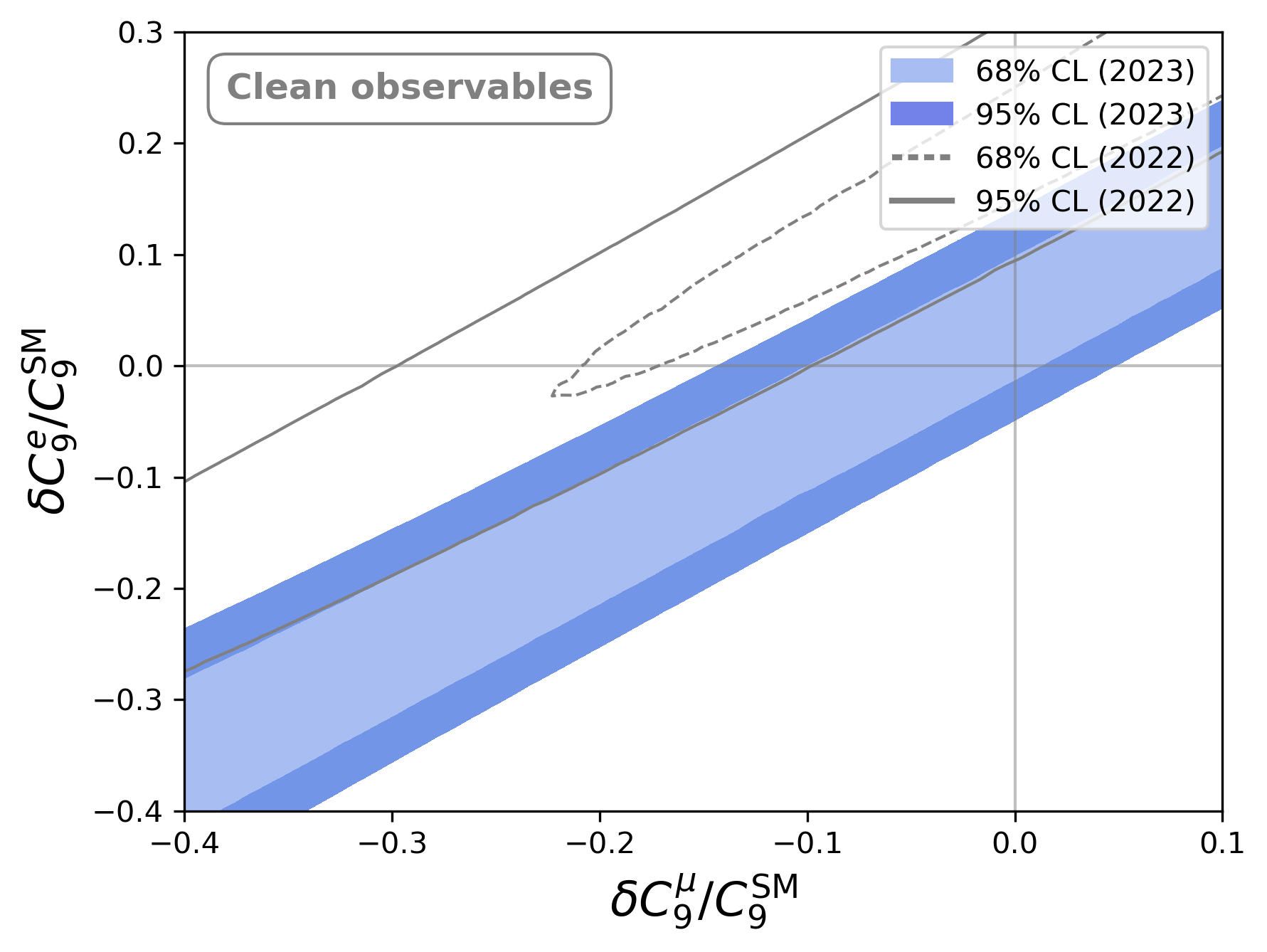}
\includegraphics[width=0.45\textwidth]{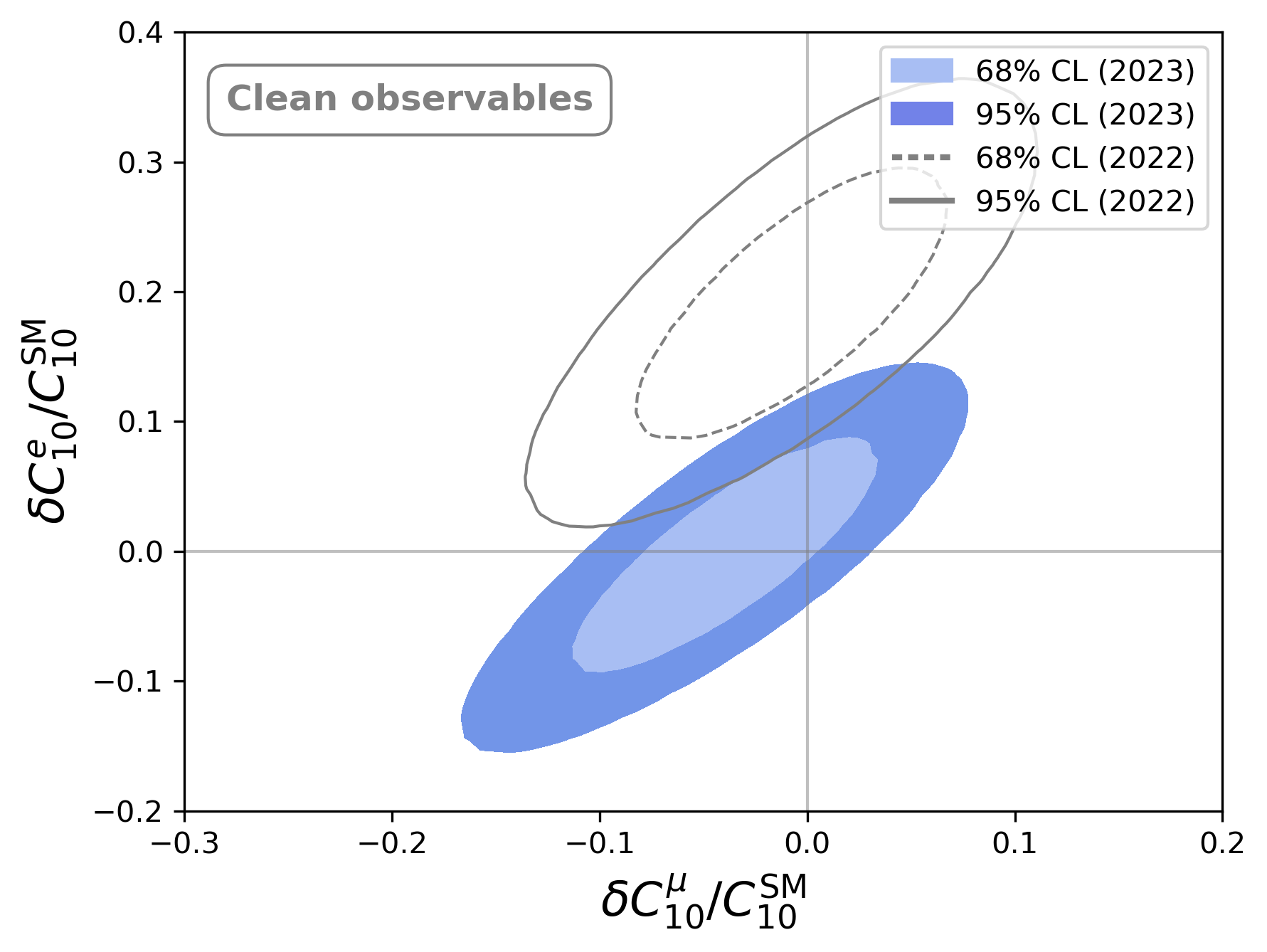}
\includegraphics[width=0.45\textwidth]{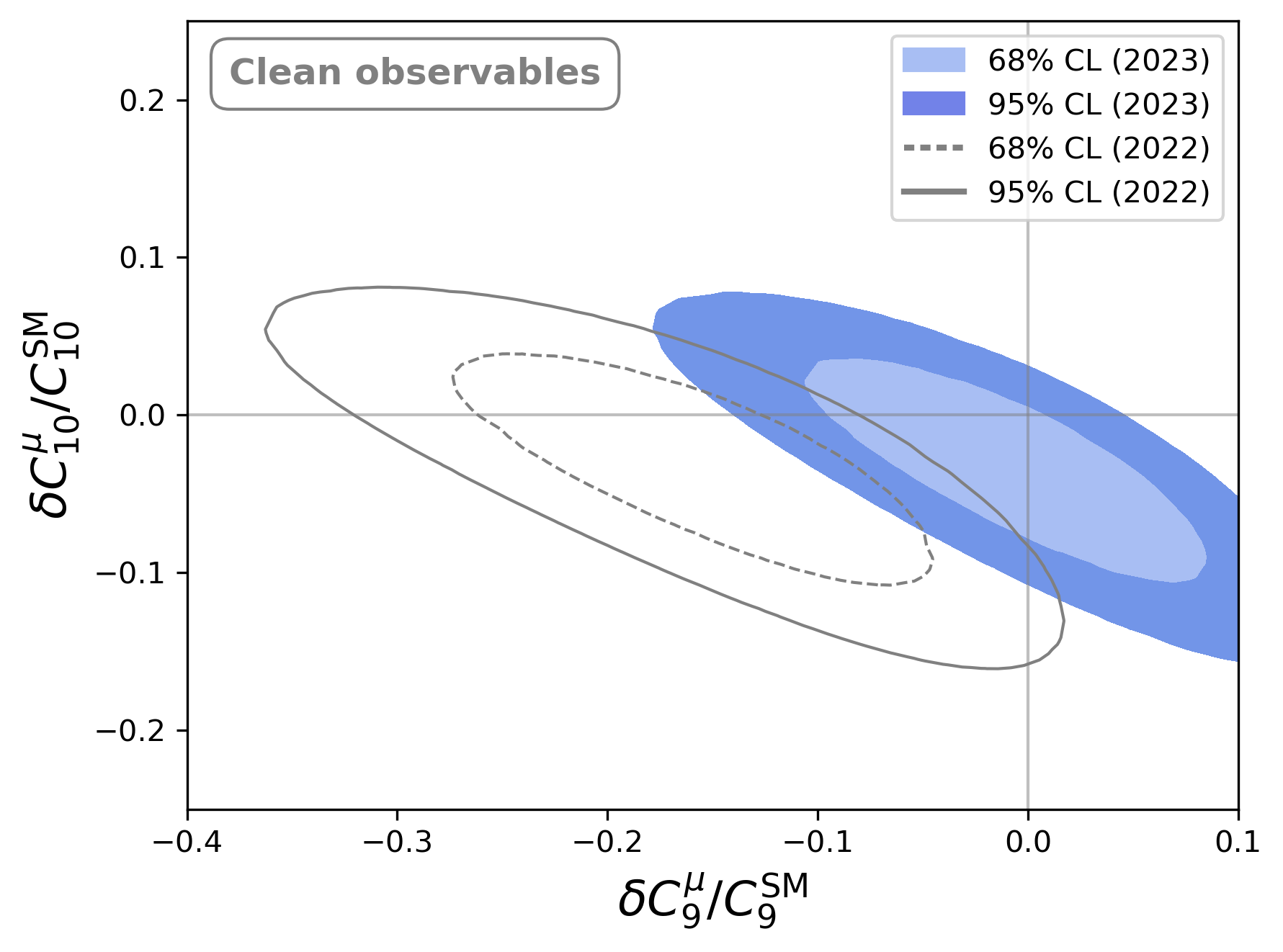}
\includegraphics[width=0.45\textwidth]{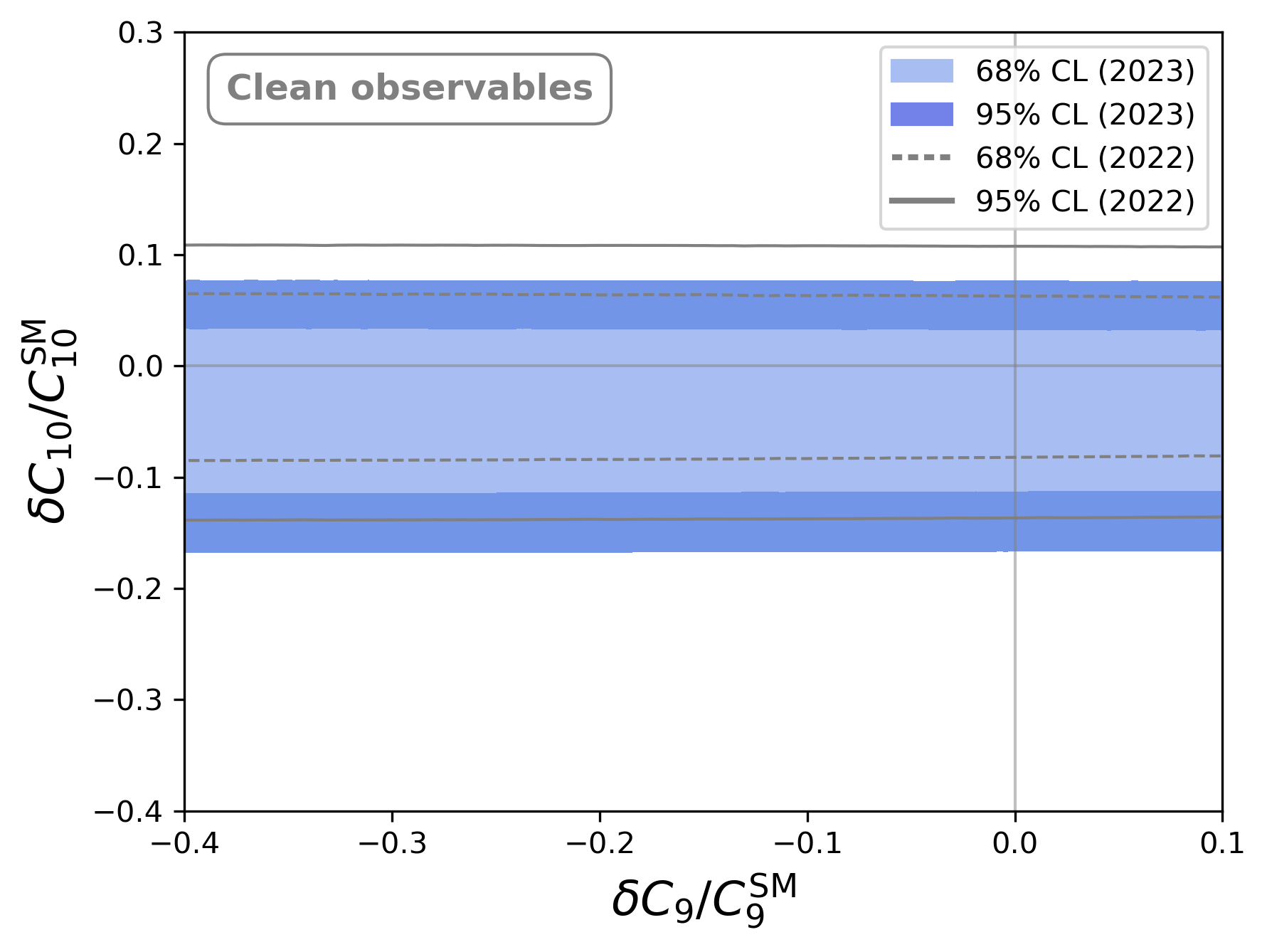}
\vspace{-0.3cm}
\caption{Two-dimensional fits to clean observables. The coloured regions correspond to the post-$R_{K^{(*)}}$ fits and the gray contours correspond to the fits prior to the recent $R_{K^{(*)}}$ update~\cite{Neshatpour:2022pvg}.
}
\label{fig:clean_2023}
\end{figure}

The corresponding two-operator fits are shown in Figure~\ref{fig:clean_2023}. The two upper plots clearly show that the new data confirms lepton universality. The 1 and $2\sigma$ regions in  the case  of $\{C_9^e,C_9^\mu\}$ and also in the case of $\{C_{10}^e,C_{10}^\mu\}$ are located around the diagonal. The favoured regions in the case of $\{C_{10}^e,C_{10}^\mu\}$ are  bounded along the diagonal because we  have included BR($B_{s,d} \to \mu^+ \mu^-$) in the fit which implies strong constraints on $C_{10}$ in general. 
The lower left plot in Figure~\ref{fig:clean_2023} shows the two-operator fit to $\{C_9^\mu,C_{10}^\mu\}$. The 1 or $2\sigma$ regions are now also grouped around the secondary diagonal and contain the SM values. Only small NP contributions are still possible after the new measurements. 
We note however that without  BR($B_{s,d} \to \mu^+ \mu^-$) in the fit, i.e. without the strong constrain on $C_{10}^\mu$ much larger values of $C_{9}^\mu$ and $C_{10}^\mu$ would be possible along the secondary diagonal. Such larger contributions are then in principle possible but due to unnatural cancellations of these two contributions in the ratios $R_K$ and $R_{K^*}$ only. 
The lower right plot is trivial. It shows that our set of clean observables  does not  constrain the universal coefficient $C_9$, but that  BR($B_{s,d} \to \mu^+ \mu^-$) constrains the universal $C_{10}$.
{The slight shift along the $C_{10}$ axis compared to pre-$R_{K^{(*)}}$ fit is due to the modified SM prediction of BR($B_s \to \mu^+\mu^-$) owing to the updated CKM inputs.}
%
%

\subsection{Fits to all \texorpdfstring{$b\to s \ell \ell $}{b->sll} data except clean observables}\label{sec:all_except_clean}
\begin{table}[th!]
\begin{center}
\setlength\extrarowheight{0pt}
\scalebox{0.85}{
\begin{tabular}{|l|r|r|c|}
\hline 
 \multicolumn{4}{|c|}{\footnotesize All observables except LFUV ratios and $B_{s,d}\to \ell^+ \ell^-$ \vspace{-0.1cm}} \\ 
 \multicolumn{4}{|c|}{{\footnotesize{\bf pre-$\boldsymbol{R_{K^{(*)}}}$ update}} \;($\chi^2_{\rm SM}=221.8$)} \\ \hline
& b.f. value & $\chi^2_{\rm min}$ & ${\rm Pull}_{\rm SM}$  \\ 
\hline \hline
$\delta C_{9} $    	& $ 	-0.95	\pm	0.13	 $ & $ 	185.1	 $ & $	6.1	\sigma	 $  \\
$\delta C_{9}^{e} $    	& $ 	0.70	\pm	0.60	 $ & $ 	220.5	 $ & $	1.1	\sigma	 $  \\
$\delta C_{9}^{\mu} $    	& $ 	-0.96	\pm	0.13	 $ & $ 	182.8	 $ & $	6.2	\sigma	 $  \\
\hline										
$\delta C_{10} $    	& $ 	0.29	\pm	0.21	 $ & $ 	219.8	 $ & $	1.4	\sigma	 $  \\
$\delta C_{10}^{e} $    	& $ 	-0.60	\pm	0.50	 $ & $ 	220.6	 $ & $	1.1	\sigma	 $  \\
$\delta C_{10}^{\mu} $    	& $ 	0.35	\pm	0.20	 $ & $ 	218.7	 $ & $	1.8	\sigma	 $  \\
\hline							          			
\hline
$\delta C_{\rm LL}^e$	& $ 	0.34	\pm	0.29	 $ & $ 	220.6	 $ & $	1.1	\sigma	 $  \\
$\delta C_{\rm LL}^\mu$    	& $ 	-0.64	\pm	0.13	 $ & $ 	195.0	 $ & $	5.2	\sigma	 $  \\
\hline										
\end{tabular}
\quad
\begin{tabular}{|l|r|r|c|}
\hline 
\multicolumn{4}{|c|}{\footnotesize { All observables except LFUV ratios and $B_{s,d}\to \ell \bar{\ell}$}} \\[-4pt]							
\multicolumn{4}{|c|}{\small{\bf { 	post-$\boldsymbol{R_{K^{(*)}}}$ update}} \; ($\chi^2_{\rm SM}=				261.6	$)} \\ \hline
& b.f. value & $\chi^2_{\rm min}$ & ${\rm Pull}_{\rm SM}$  \\							
\hline \hline								
$\delta C_{9} $    	& $ 	-0.97	\pm	0.13	 $ & $ 	221.9	 $ & $	{6.3	\sigma}	 $  \\
$\delta C_{9}^{e} $    	& $ 	0.70	\pm	0.60	 $ & $ 	260.4	 $ & $	1.1	\sigma	 $  \\
$\delta C_{9}^{\mu} $    	& $ 	-0.98	\pm	0.13	 $ & $ 	219.7	 $ & ${	6.5	\sigma}	 $  \\
\hline										
$\delta C_{10} $    	& $ 	0.36	\pm	0.20	 $ & $ 	258.3	 $ & $	1.8	\sigma	 $  \\
$\delta C_{10}^{e} $    	& $ 	-0.50	\pm	0.50	 $ & $ 	260.5	 $ & $	1.0	\sigma	 $  \\
$\delta C_{10}^{\mu} $    	& $ 	0.41	\pm	0.20	 $ & $ 	257.0	 $ & $	2.1	\sigma	 $  \\
\hline							          			
\hline										
$\delta C_{\rm LL}^e$	& $ 	0.31	\pm	0.28	 $ & $ 	260.4	 $ & $	1.1	\sigma	 $  \\
$\delta C_{\rm LL}^\mu$    	& $ 	-0.65	\pm	0.12	 $ & $ 	231.7	 $ & $	{ 5.5	\sigma}	 $  \\
\hline										
\end{tabular}
} 
\caption{One operator fits for all except clean  observables before (left)  and also after (right) the LHCb-update of $R_{K^{(*)}}$.
\label{tab:2022_2023_1D_all_but_clean}} 
\end{center} 
\end{table}
In Table~\ref{tab:2022_2023_1D_all_but_clean} we show the one-parameter fits to the rest of the $b \to s$ observables - excluding the clean observables discussed before. These fits are of course almost unchanged compared to the situation before the new measurements of $R_K$ and $R_{K^*}$.  The slight differences in the NP significance are  due to the new measurements by CMS and also update of the CKM parameters. 
But the comparison of the one-operator fits to the clean observables in Table~\ref{tab:2022_2023_1D_clean}  and of  those to  the remaining $b \to s$ observables in Table~\ref{tab:2022_2023_1D_all_but_clean} shows for the non-universal Wilson coefficients $C_9^\mu$ and $C_{LL}^\mu$ no longer any consistency, which means that the remaining large tensions in the rest of the $b \to s$ observables, in particular in the angular observables and in the branching ratios, should be described with lepton-universal operators, only small deviations from the lepton universality are allowed. {Let us emphasise that the NP significances given in  
Table~\ref{tab:2022_2023_1D_all_but_clean}
are based on the assumption of $10\%$ power corrections to  the angular observables and branching ratios. }

%
\subsection{Fits to all \texorpdfstring{$b\to s\ell\ell$}{b->sll} observables}\label{sec:all_obs}

This brings us to the fits to all $b \to s$ observables where we now use lepton-universal operators only -- {assuming again $10\%$ power corrections for  the angular observables and branching ratios.} The results are  given in Table~\ref{tab:2022_2023_1D_all} where we can see that the favoured universal coefficient is $C_9$  in order to explain  the tensions in the angular observables and branching ratios.  In principle, $C_9^\mu$ and $C_{LL}^\mu$ can explain the tensions but these new physics contributions would not be compatible with the constraints induced by the clean observables as we showed above.  In Table~\ref{tab:2022_2023_1Dchiral_all} one operator fits using chiral universal coefficients~\footnote{We use the standard notation:  $C_{XY}$ where $X$ denotes the chirality of the quark current and $Y$ of the lepton one.
Assuming left-handed leptons only, we have $C_{LL} \equiv  C_9 = - C_{10}$ and $C_{RL} \equiv  C^{\prime}_9 = - C^{\prime}_{10}$\,, for right-handed leptons  $C_{RR} \equiv C^{\prime}_9 = C^{\prime}_{10}$ and    $C_{LR} \equiv {C_9} = C_{10}\,.$} are shown.   One finds a rather large NP significance for the fits to $C_{LL}$ and $C_{LR}$, i.e. for left-handed quark currents.

%
\begin{table}[th!]
\begin{center}
\setlength\extrarowheight{3pt}
\hspace*{-1.cm}
\scalebox{0.97}{
\begin{tabular}{|l|r|r|c|}
\hline 
\multicolumn{4}{|c|}{All observables} \\[-4pt]										
\multicolumn{4}{|c|}{\small{\bf pre-$\boldsymbol{R_{K^{(*)}}}$ update} \; ($\chi^2_{\rm SM}=				253.5	$)} \\ \hline			
& b.f. value & $\chi^2_{\rm min}$ & ${\rm Pull}_{\rm SM}$  \\										
\hline \hline										
$\delta C_7$	& $ 	-0.02	\pm	0.01	 $ & $ 	248.7	 $ & $	2.2	\sigma	 $  \\
\hline										
$\delta C_{Q_{1}} $    	& $ 	-0.05	\pm	0.02	 $ & $ 	252.3	 $ & $	1.1	\sigma	 $  \\
$\delta C_{Q_{2}} $    	& $ 	-0.01	\pm	0.01	 $ & $ 	252.4	 $ & $	1.0	\sigma	 $  \\
\hline										
$\delta C_{9} $    	& $ 	-0.95	\pm	0.13	 $ & $ 	215.8	 $ & $	6.1	\sigma	 $  \\
$\delta C_{10} $    	& $ 	0.08	\pm	0.16	 $ & $ 	253.2	 $ & $	0.5	\sigma	 $  \\
\hline							          			
\end{tabular}
\begin{tabular}{|l|r|r|c|}
\hline 
\multicolumn{4}{|c|}{All observables} \\[-4pt]										
\multicolumn{4}{|c|}{\small{\bf { post-$\boldsymbol{R_{K^{(*)}}}$ update}} \; ($\chi^2_{\rm SM}=				271.0	$)} \\ \hline			
& b.f. value & $\chi^2_{\rm min}$ & ${\rm Pull}_{\rm SM}$  \\										
\hline \hline										
$\delta C_7$	& $ 	-0.02	\pm	0.01	 $ & $ 	267.2	 $ & $	1.9	\sigma	 $  \\
\hline 
$\delta C_{Q_{1}} $    	& $ 	-0.04	\pm	0.03	 $ & $ 	270.3	 $ & $	0.8	\sigma	 $  \\
$\delta C_{Q_{2}} $    	& $ 	-0.01	\pm	0.01	 $ & $ 	270.4	 $ & $	0.8	\sigma	 $  \\
\hline										
$\delta C_{9} $    	& $ 	-0.96	\pm	0.13	 $ & $ 	230.7	 $ & $	{ 6.3	\sigma}	 $  \\
$\delta C_{10} $    	& $ 	0.15	\pm	0.15	 $ & $ 	270.0	 $ & $	1.0	\sigma	 $  \\
\hline							          			
\end{tabular}
}
\caption{One operator NP fits to all $b\to s \ell \ell$ observables  before and after the update of $R_{K^{(*)}}$ by the LHCb collaboration. 
\label{tab:2022_2023_1D_all}
} 
\end{center} 
\end{table}
%

\begin{table}[th!]
\begin{center}
\setlength\extrarowheight{3pt}
\hspace*{-1.cm}
\scalebox{0.97}{
\begin{tabular}{|l|r|r|c|}
\hline 
\multicolumn{4}{|c|}{ All observables} \\[-4pt]										
\multicolumn{4}{|c|}{\small{\bf {	post-$\boldsymbol{R_{K^{(*)}}}$ update	}} \; ($\chi^2_{\rm SM}=				271.0	$)} \\ \hline			
& b.f. value & $\chi^2_{\rm min}$ & ${\rm Pull}_{\rm SM}$  \\										
\hline \hline										
$\delta C_{\rm LL}$	& $ 	-0.54	\pm	0.12	 $ & $ 	249.1	 $ & $	{ 4.7	\sigma}	 $  \\
\hline										
$\delta C_{\rm LR}$	& $ 	-0.42	\pm	0.10	 $ & $ 	257.4	 $ & $	{ 3.7	\sigma}	 $  \\
\hline										
$\delta C_{\rm RL}$	& $ 	0.00	\pm	0.08	 $ & $ 	268.8	 $ & $	1.5	\sigma	 $  \\
\hline										
$\delta C_{\rm RR}$	& $ 	0.21	\pm	0.13	 $ & $ 	268.1	 $ & $	1.7	\sigma	 $  \\
\hline
\end{tabular}
}
\caption{One operator fits to all $b\to s \ell \ell$ observables in the chiral basis.
\label{tab:2022_2023_1Dchiral_all}} 
\end{center} 
\end{table}
%
%

\begin{table}[th!]
\begin{center}
\setlength\extrarowheight{3pt}
\hspace*{-1.cm}
\scalebox{0.9}{
\begin{tabular}{|c|c|c|c|}
\hline
\multicolumn{4}{|c|}{All observables} \\[-4pt]													
 \multicolumn{4}{|c|}{\small{\bf   	post-$\boldsymbol{R_{K^{(*)}}}$ update	} \;  ($\chi^2_{\rm SM}=		271.0	$)} \\ \hline								
& b.f. value & $\chi^2_{\rm min}$ & ${\rm Pull}_{\rm SM}$ \\ 	 												
\hline \hline													
$\{ \delta C_{9}^{\mu} , \delta C_{9}^{e} \}$	& $\{	-0.96	\pm	0.13	,\,	-0.74	\pm	0.21	\}$ & $	228.8	$ & $	6.2	\sigma$ \\
$\{ \delta C_{10}^{\mu} , \delta C_{10}^{e} \}$	& $\{	0.15	\pm	0.15	,\,	-0.03	\pm	0.21	\}$ & $	268.3	$ & $	1.1	\sigma$ \\
$\{ \delta C_{9}^{\mu} , \delta C_{10}^{\mu} \}$	& $\{	-0.78	\pm	0.12	,\,	-0.19	\pm	0.10	\}$ & $	237.2	$ & $	5.5	\sigma$ \\
$\{ \delta C_{9} , \delta C_{10} \}$	& $\{	-0.97	\pm	0.13	,\,	0.09	\pm	0.15	\}$ & $	230.3	$ & $	6.0	\sigma$ \\
\hline 
\end{tabular}
}
\caption{Two operator NP fits to all observables (post-$R_{K^{(*)}}$ update).
The corresponding plots are given in Fig.~\ref{fig:all_2023}. 
\label{tab:2022_2023_2D_all_CMS}} 
\end{center} 
\end{table}
%
%
%
%
%
%
\begin{figure}[hbt!]
\centering
\includegraphics[width=0.45\textwidth]{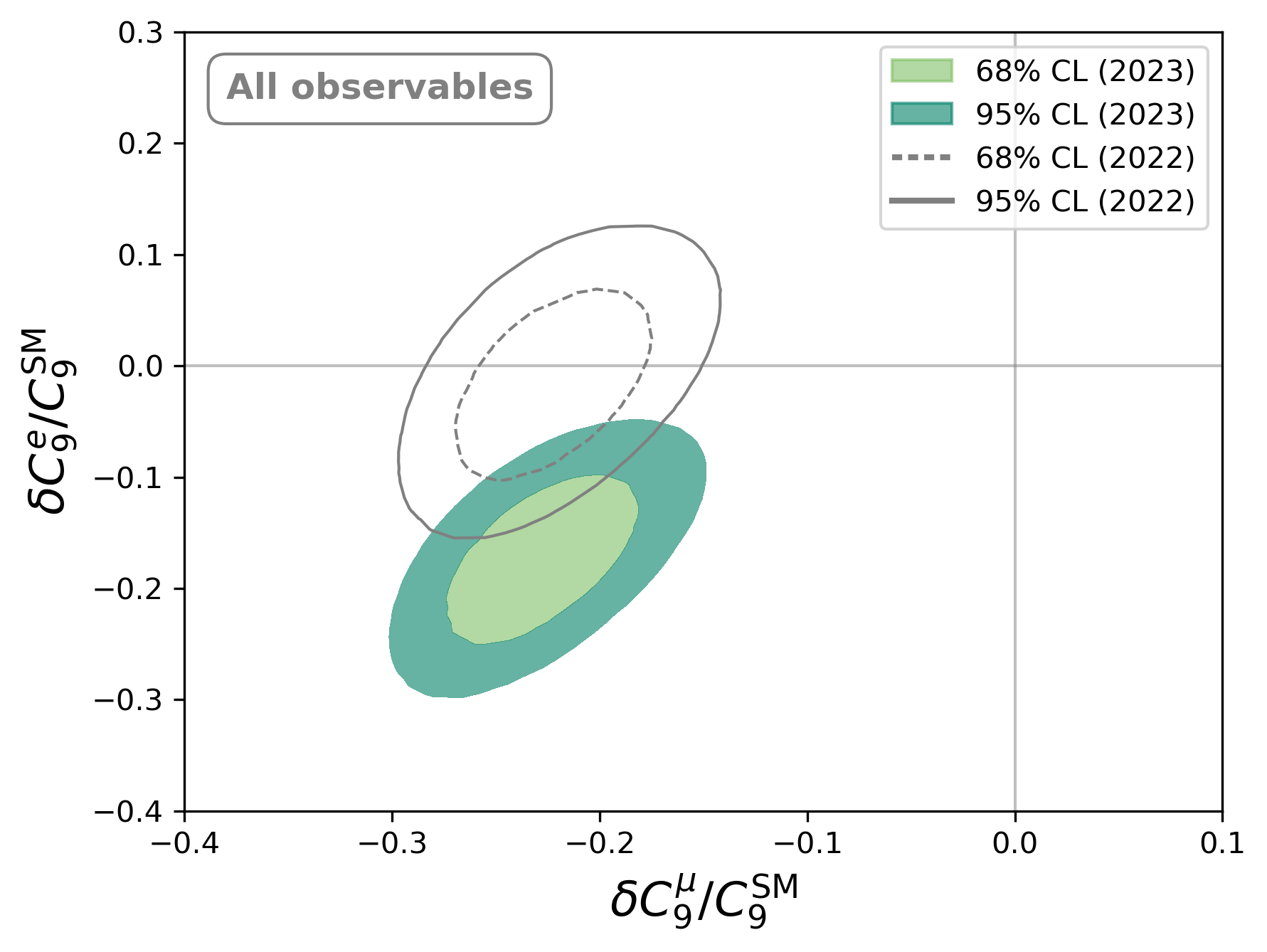}
\includegraphics[width=0.45\textwidth]{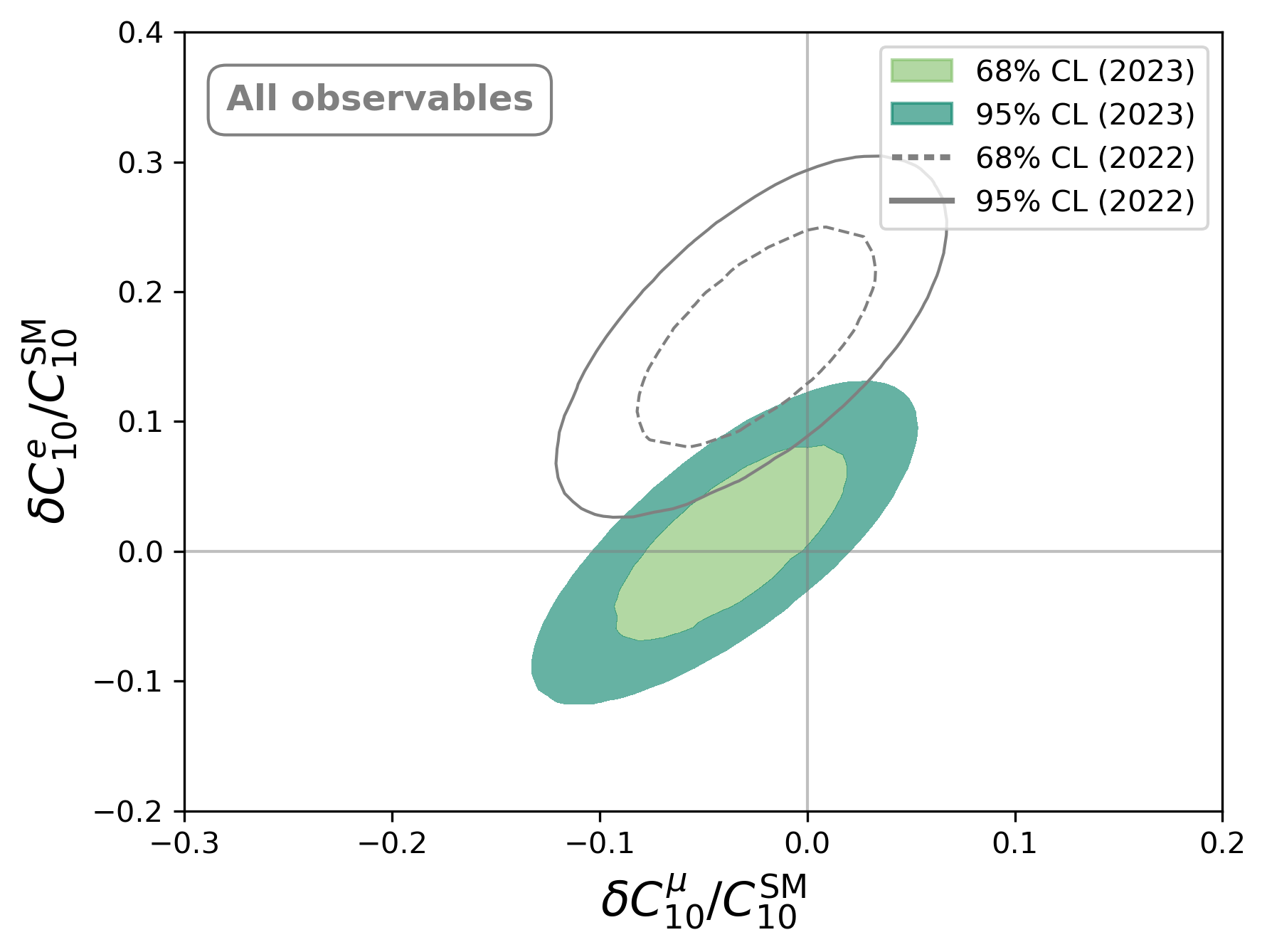}
\includegraphics[width=0.45\textwidth]{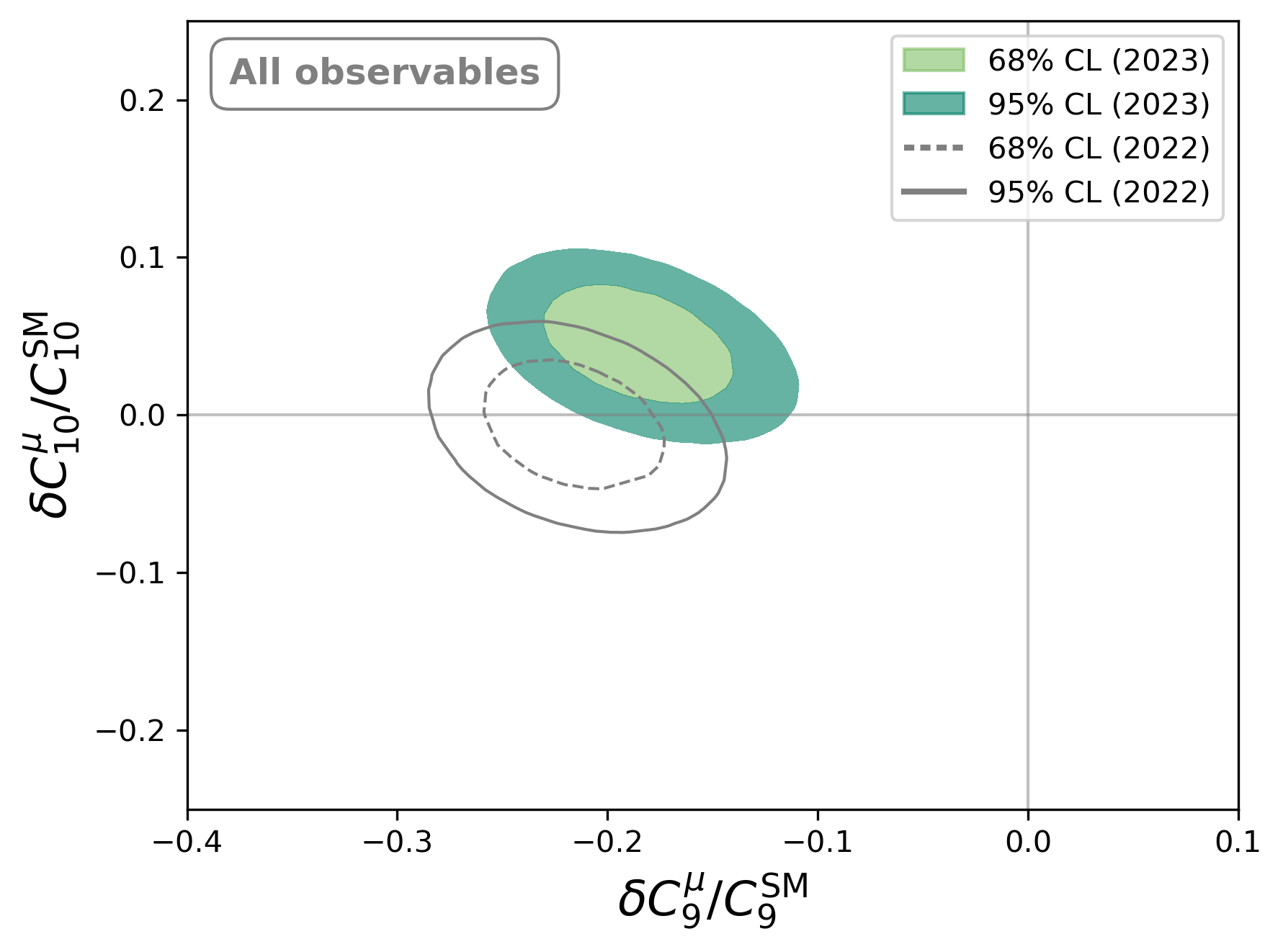}
\includegraphics[width=0.45\textwidth]{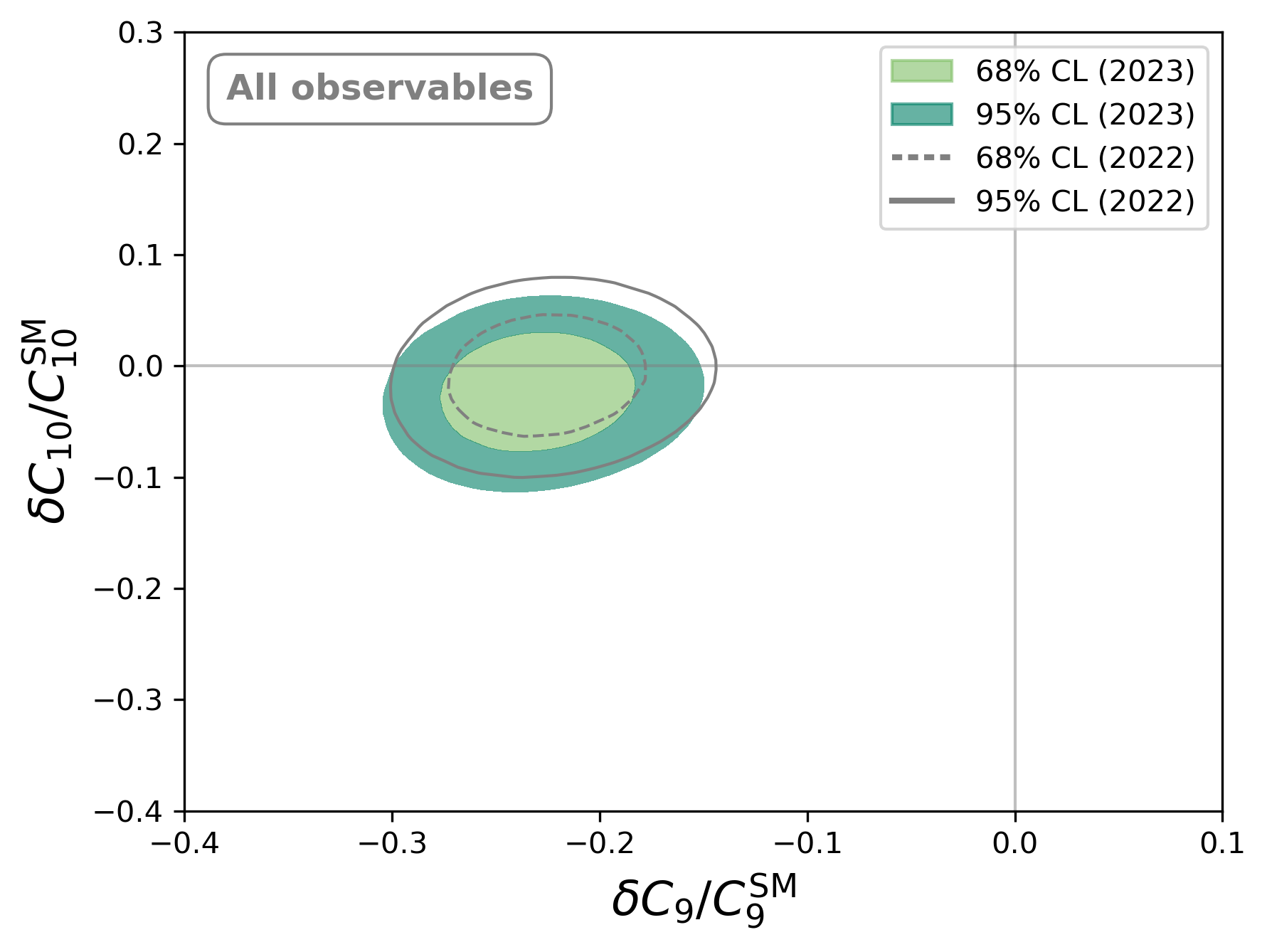}
\vspace{-0.3cm}
\caption{Two-dimensional fits to all observables {with the best-fit point} given in Table~\ref{tab:2022_2023_2D_all_CMS}. }
\label{fig:all_2023}
\end{figure}
In addition, we present the two-dimensional fit results in Figure~\ref{fig:all_2023}. The lower right plot in the $\{C_9, C_{10}\}$ plane is the crucial one. It shows that not the universal coefficient $C_{10}$ but $C_9$  explains the present anomalies best. This is also a consequence of the  $C_{10}$ dependence  of the $B_{s} \to \mu^+\mu^-$ branching ratio which is SM-like.  
The two-operator fits in the upper row, $\{  C_{9}^{\mu} ,  C_{9}^{e} \}$ and    $\{  C_{10}^{\mu} ,  C_{10}^{e} \}$, essentially are again consequences of lepton-{flavour} universality. In both plots, the 1 and 2$\sigma$ ranges have move to the diagonal and have become thinner compared to the ones of the pre-$R_{K^{(*)}}$ measurements.  Moreover, the $1\sigma$ range of $\{  C_{10}^{\mu} ,  C_{10}^{e} \}$ fit includes the SM values. It becomes clear that these two-operator plots  essentiallly reproduce the one-operator fits to the corresponding  universal $C_ 9$ and $C_{10}$.  Also the NP significance is similar as one can read off from Tables~\ref{tab:2022_2023_1D_all} and~\ref{tab:2022_2023_2D_all_CMS}.

The plot in the lower row on the left shows the two-operator fit to  $\{ C_{9}^{\mu} , C_{10}^{\mu} \}$.  Compared to the pre-$R_{K^{(*)}}$ update, the 1 or 2$\sigma$ ranges  now move in the direction of the second diagonal to allow a partial  compensation of the $C_9^\mu$ 
and the $C_{10}^\mu$ contributions within the $R_{K^{(*)}}$ ratios.
Because of this unnatural compensation, this specific two-operator fit should be considered critical.
In comparison with the corresponding plot  in Figure~\ref{fig:clean_2023} with the fits to the clean observables, one needs now a larger $C_9^\mu$ for the explanation of the present tensions
which again indicates the present measurements are best described by flavour-universal operators.

As one can read off from Table~\ref{tab:2022_2023_2D_all_CMS}, all two-operator fits discussed have a large NP significance up to 6$\sigma$ besides the case $\{ C_{10}^{\mu} , C_{10}^{e} \}$.

\begin{figure}[t!]
\centering
\includegraphics[width=0.49\textwidth]{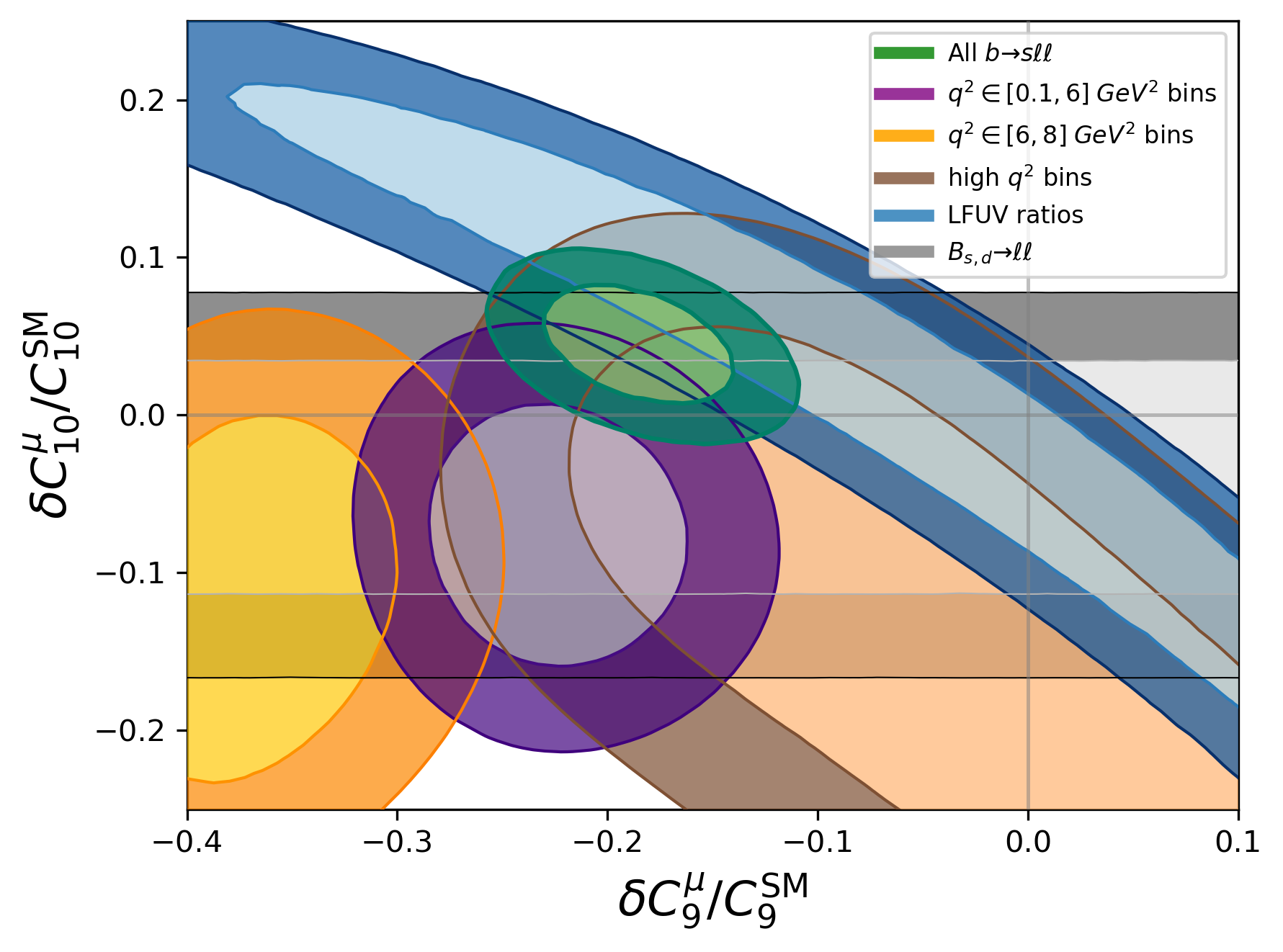}
\includegraphics[width=0.49\textwidth]{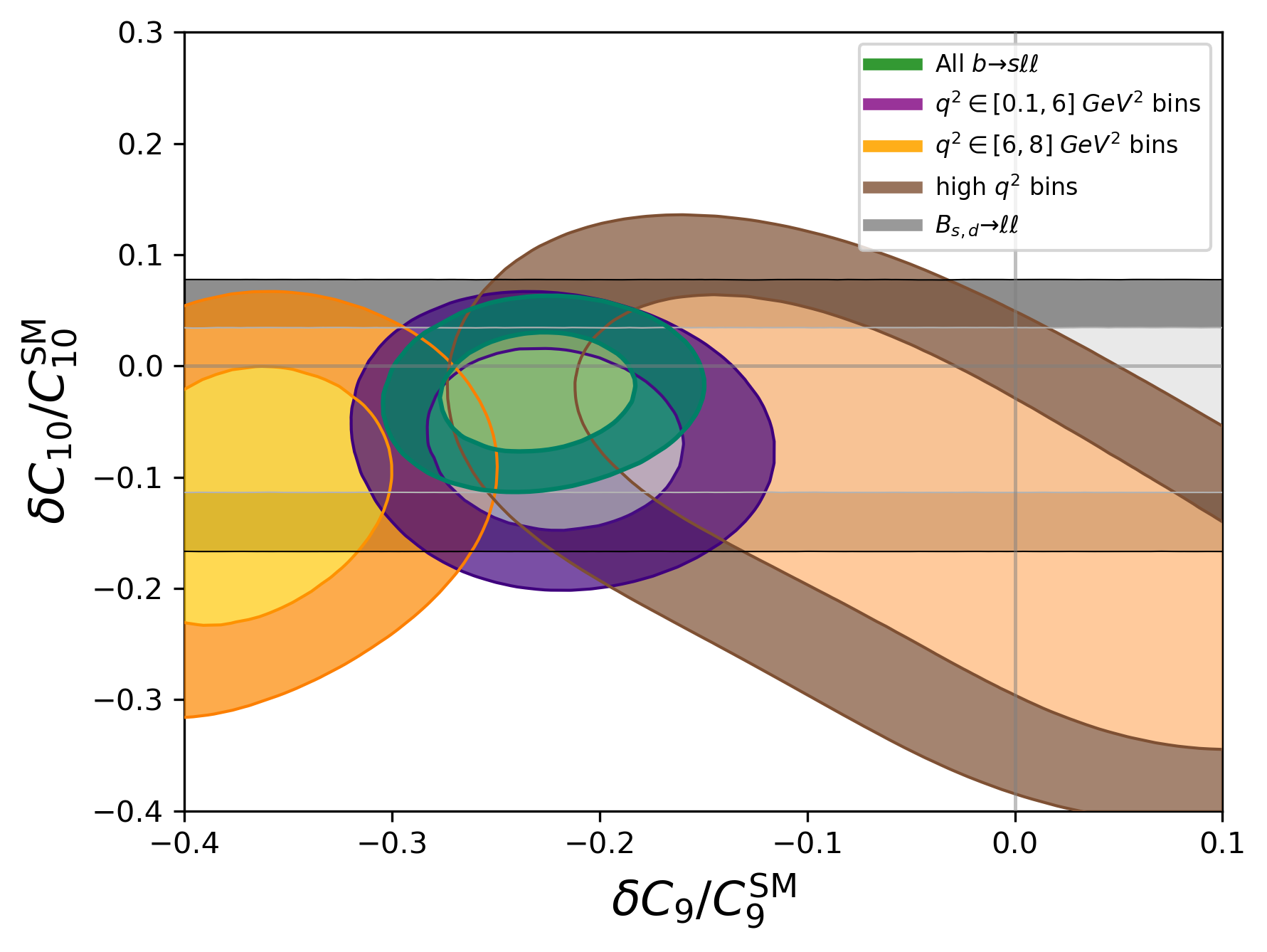}
\vspace{-0.3cm}
\caption{Two-dimensional fits to all observables in green. Where relevant, the impact of the $b\to s \ell \ell$ observables for the low $q^2$ bins up to 6 GeV$^2$, for the $[6,8]$ GeV$^2$ bin and for the high $q^2$ bins as well as the bounds from the lepton flavour universality violating ratios and  $B_{s,d}\to \ell^+ \ell^-$ are shown separately with the lighter (darker) shade indicating the 68\% (95\%) confidence level region.} 
\label{fig:all_separated_2023}
\end{figure}

Next, we will have  a closer look at the two two-operator fits 
to $\{ C_{9}^{\mu},C_{10}^{\mu} \}$ and $\{ C_{9} , C_{10}\}$. 
We consider the bounds of the $R_{K^{(*)}}$  ratios  separately from the ones induced by the  $B_{s,d} \to \mu^+\mu^-$ branching ratios. Likewise, in the case of the remaining $b \to s\ell^+\ell^-$ observables, we examine the impact of the low-$q^2$  and  the high-$q^2$ observables separately.   
Since the validity of SCET in the low-$q^2$ bin $[6,8]$ GeV$^2$ (near the $J/\psi$ resonance) is questionable, we separate this  bin from the other low-$q^2$ bins up to $6$  GeV$^2$. 

In Figure~\ref{fig:all_separated_2023}  the 
two-operator fits have been dissected in order to show the impact that each of these different sets of observables have on the overall fit.
In the plot on the right hand side of Figure~\ref{fig:all_separated_2023}   the $\{ C_{9} ,C_{10}\}$
two-operator fit has been shown, where the brown contours show the 1 and 2$\sigma$ regions of the high-$q^2$  observables.  It can be seen that they are compatible with the SM values with comparatively large uncertainties. The tensions in the angular observables and the branching rations obviously have their main origin in the low-$q^2$ observables as can be seen from the purple contours. It is well-known that the high-$q^2$ observables have a weak dependence on the Wilson coefficients, which implies a low sensitivity to NP.~\footnote{In principle the high-$q^2$ observables are theoretically cleaner. There is a
local operator product expansion (OPE) to describe power corrections (see i.e. Refs.~\cite{Bobeth:2010wg,Beylich:2011aq}).}  The yellow contours show that the inclusion of the highest low-$q^2$ bin from 6 to 8 
GeV$^2$ in the fit massively increases the NP significance.  
However, it could be that this large effect just indicates that SCET is no longer valid in this range. Finally, the $B_{s,d} \to \mu^+\mu^-$ branching ratios lead to the grey contours which just bound the Wilson coefficient $C_{10}$.

In the plot on the left  hand side  of Figure~\ref{fig:all_separated_2023} we look at the bounds on $\{ C_{9}^{\mu} , C_{10}^{\mu} \}$.   The blue 1 and 2$\sigma$ regions show the bounds  generated by the ratios  $R_{K^{(*)}}$. This can be compared to the lower right plot of Figure~\ref{fig:clean_2023} where the bound from the ratios together with BR($B_{s,d} \to \mu^+\mu^-$) was shown. 
One realises that now much larger values of $C_9^\mu$ and also of $C_{10}^\mu$ are allowed, but this is possible due to an unnatural compensation between  the $C_9^\mu$ and the  $C_{10}^\mu$ contributions in the ratios which makes the $\{ C_{9}^{\mu} , C_{10}^{\mu} \}$  fit problematic, as already mentioned above. 
The  $B_{s,d} \to  \mu^+\mu^-$ branching ratios alone bound  $C_{10}^\mu$ to smaller values again, as can be seen from the grey  contours.


\section{Global analyses}\label{sec:global}
In order to present the global analysis, we provide multi-dimensional fits considering only universal operators, which may be more realistic than assuming one- or two-operator fits, since it is unlikely that a complete NP scenario would affect only one parameter while leaving the others unchanged. 
We therefore consider a fit varying simultaneously all the relevant 12 lepton-flavour universal Wilson coefficients. This multi-dimensional fit also avoids the look-elsewhere effect (LEE), which can occur when making a selected choice of observables or when assuming a subset of specific new physics directions.  The results are presented in Table~\ref{tab:12D_2023}. As can be seen, most primed coefficients (with right-handed quark currents) are only loosely constrained with the currently available data.
\begin{table}[t!]
\begin{center}
\setlength\extrarowheight{3pt}
\scalebox{0.9}{
\begin{tabular}{|c|c|c|c|}
\hline																
\multicolumn{4}{|c|}{All observables  with $\chi^2_{\rm SM}=	 	271.0			$} \\											
\multicolumn{4}{|c|}{\small{\bf 	post-$\boldsymbol{R_{K^{(*)}}}$ update}\; 		$(\chi^2_{\rm min}=	 	222.5	;\; {\rm Pull}_{\rm SM}=	4.7				\sigma$)} \\					
\hline \hline																
\multicolumn{2}{|c|}{$\delta C_7$} &  \multicolumn{2}{c|}{$\delta C_8$}\\																
\multicolumn{2}{|c|}{$	0.07	\pm	0.03	$} & \multicolumn{2}{c|}{$	-0.70	\pm	0.50	$}\\								
\hline																
\multicolumn{2}{|c|}{$\delta C_7^\prime$} &  \multicolumn{2}{c|}{$\delta C_8^\prime$}\\																
\multicolumn{2}{|c|}{$	-0.01	\pm	0.01	$} & \multicolumn{2}{c|}{$	-0.50	\pm	1.20	$}\\								
\hline																
$\delta C_{9}$ & $\delta C_{9}^{\prime}$ & $\delta C_{10}$ & $\delta C_{10}^{\prime}$ \\																
$	-1.18	\pm	0.19	$ & $	0.06	\pm	0.31	$ & $	0.23	\pm	0.20	$ & $	-0.05	\pm	0.19	$ \\
\hline\hline																
$C_{Q_{1}}$ & $C_{Q_{1}}^{\prime}$ & $C_{Q_{2}}$ & $C_{Q_{2}}^{\prime}$ \\																
$	-0.30	\pm	0.14	$ & $	-0.18	\pm	0.14	$ & $	0.01	\pm	0.02	$ & $	-0.03	\pm	0.07	$ \\
\hline																
\end{tabular}
} 
\caption{The 12-dimensional (lepton flavour universal) fit to all observables.
\label{tab:12D_2023}} 
\end{center} 
\end{table}
%
%
In Table~\ref{tab:Wilks_LFU_2023} we compare the significance of different NP fits (all lepton-flavour universal) compared to the SM and to each other considering the Wilks' theorem~\cite{Wilks:1938dza}. Since the NP scenarios in Table~\ref{tab:Wilks_LFU_2023} are nested in the model of the next row, we can calculate $p$-values using Wilks' theorem. The difference in $\chi^2$ between the two models is itself a $\chi^2$-distribution with a number of degrees of freedom equal to the difference in the number of parameters. The $p$-value therefore indicates the significance of the new parameters added.  We have then converted these p-values to sigmas. From Table~\ref{tab:Wilks_LFU_2023}, it is clear that the main coefficient explaining the measured tensions in $b\to s$ decays is $C_9$ and beyond that adding further degrees of freedom does not improve the fit significantly.
Thus, also the Wilks' test confirms the crucial role of $C_9$ for the explanation of the anomalies in the angular observables and branching ratios.
\begin{table}[h]
\begin{center}
\scalebox{0.95}{
\begin{tabular}{|c|c|c|c|c|}\hline
\multicolumn{5}{|c|}{All observables\;	 {\small{\bf (post-$\boldsymbol{R_{K^{(*)}}}$ update)}}} \\				
\hline \hline									
Set of WC & param. & $\chi^2_{\rm min}$ & Pull$_{\rm SM}$ & Improvement\\ \hline									
SM                                          	&	0	&	271.0	& $	-	$ & $	-	$ \\
$C_9$                                 	&	1	&	230.7	& $	6.3	\sigma$ & $	6.3	\sigma$\\
$C_9, C_{10}$                   	&	2	&	230.3	& $	6.0	\sigma$ & $	0.6	\sigma$\\
$C_7,C_8,C_9,C_{10}$	&	4	&	225.3	& $	5.9	\sigma$ & $	1.7	\sigma$\\
$C_7,C_8,C_9,C_{10},C_{Q_1},C_{Q_2}$    	&	6	&	224.7	& $	5.6	\sigma$ & $	0.3	\sigma$\\
All WC (incl. primed)                       	&	12	&	222.5	& $	4.7	\sigma$ & $	0.1	\sigma$\\
\hline
\end{tabular}
}
\caption{\small Pull$_{\rm SM}$ of $1,2,4,6$ and 12 dimensional fit. 
The last row includes all Wilson coefficients including the chirality-flipped primed coefficients. The last column indicates the significance of the improvement of the fit compared to the previous row.
\label{tab:Wilks_LFU_2023}}\vspace*{-0.6cm}
\end{center} 
\end{table}

\section{Summary}\label{sec:summary}
{
In light of the recent LHCb measurement of $R_K$ and $R_{K^*}$ which  is in agreement with the Standard Model (SM) prediction, we have analysed the current status of $b \to s$ semileptonic decays, including this new measurement, as well as the very recent measurement of $R_K$ and BR($B^+ \to K^+ \mu^+ \mu^-$) by the CMS collaboration. We have also updated the CKM parameters to the PDG2022 values.

The clean observables $R_K$, $R_{K^*}$, and BR($B_s \to \mu^+ \mu^-$) are now all in good agreement with the SM. The ratios constrain new physics contributions in $b \to s \ell^+ \ell^-$ decays to be lepton flavour universal, with room for only small universality violating contributions while BR($B_s \to \mu^+ \mu^-$) constrains new physics contributions in the axial Wilson coefficient $C_{10}$.
Furthermore, we showed that although the two-dimensional fit $\{C_9^\mu,C_{10}^\mu\}$ (with $C_{9}^e$ and $C_{10}^e$ kept to their SM values) indicates preference for NP in $C_9^\mu$ and to a lesser degree in $C_{10}^\mu$, this two-operator fit should be viewed critically because it gives a LFUV solution  which is at odds with the recent $R_{K^{(*)}}$ measurements.

However, the tensions in the angular observables and branching ratios are untouched by the new LHCb measurements. These tensions are best explained by a lepton flavour universal NP in the Wilson coefficient $C_9$, which is mostly due to the low-$q^2$ observables, especially from the $[6-8]\;\text{GeV}^2$ bin, keeping in mind that this latter on the one hand is more sensitive to $C_9$ contributions, and on the other hand more prone to being contaminated by charm-loop contributions.
{Moreover, as shown via the Wilks' test, new physics contributions in $C_9$ is the main scenario explaining the measured tensions in $b\to s$ decays and there is no significant improvement in the fit when considering more complex models with additional degrees of freedom.}
}

\section*{Acknowledgement} 
The authors are grateful to  P.~Owen for useful discussions.
TH is  supported by  the  Cluster  of  Excellence  ``Precision  Physics,  Fundamental Interactions, and Structure of Matter" (PRISMA$^+$ EXC 2118/1) funded by the German Research Foundation (DFG) within the German Excellence Strategy (Project ID 39083149).
SN was supported in part by the INFN research initiative Exploring New
Physics (ENP).

\bibliography{Brefs}

\end{document}